# SwinGS: Sliding Window Gaussian Splatting for Volumetric Video Streaming with Arbitrary Length

Bangya Liu
University of Wisconsin-Madison
bangya@cs.wisc.edu

Suman Banerjee
University of Wisconsin-Madison
suman@cs.wisc.edu

## Abstract

Recent advances in 3D Gaussian Splatting (3DGS) have garnered significant attention in computer vision and computer graphics due to its high rendering speed and remarkable quality. While extant research has endeavored to extend the application of 3DGS from static to dynamic scenes, such efforts have been consistently impeded by excessive model sizes, constraints on video duration, and content deviation. These limitations significantly compromise the streamability of dynamic 3D Gaussian models, thereby restricting their utility in volumetric video streaming.

This paper introduces **SwinGS**, a streaming-friendly paradigm representing volumetric video as a per-frame-update 3D Gaussian model that could be easily scaled to arbitrary video length. Specifically, we incorporate a sliding-window based continuous training during the train stage as well as a straightforward rendering at client side. We implement a prototype of **SwinGS** and demonstrate its streamability with DyNeRF dataset. Additionally, we develop an interactive WebGL viewer enabling real-time volumetric video playback on most devices with modern browsers, including smartphones and tablets. Experimental results show that **SwinGS** reduces transmission costs by 83.6% compared to previous work and could be easily scaled to volumetric videos with arbitrary length with no increasing of required GPU resources.

## 1. Introduction

Volumetric video, also known as free view video (FVV), represents a revolutionary media format that enables viewers to experience content as if physically present in the scene. Unlike traditional video captured from a single perspective, volumetric video encapsulates the depth, shape, and motion of objects as well as people within a scene. This 3D representation can be viewed from any position or perspective in virtual reality (VR), augmented reality (AR), or on flat screens with user interaction. Beyond entertainment, volumetric video plays crucial roles in autonomous vehicles [40], robotics vision, and teleoperation [30].

Historically, volumetric video has relied on point clouds [9, 15] and meshes as foundational elements. However, these approaches have struggled to balance video quality with storage and bandwidth efficiency. Recent advances in computer graphics have introduced a new family of 3D scene representations: neural rendering. This includes Neural Radiance Fields (NeRF) [25] and the emerging 3D Gaussian Splatting (3DGS) [13]. While NeRF achieves superior rendering quality with compact storage, it suffers from intensive computational costs due to its sampling process, resulting in low frame rates. The 3D Gaussian model, a "volumetric and differentiable" variant of point clouds, has emerged as a promising alternative.

Industry demos of 3DGS, including Teleport from Varjo [33] and Gracia [1], have garnered significant attention upon release. More recently, static 3D Gaussian Splatting has been showcased on the Pico XR headset [2] and Apple Vision Pro via MetalSplatter [3], offering impressive immersive experiences.

Recent research works [6, 10, 12, 18, 23, 31, 34, 36–38] have demonstrated the potential of 3DGS in representing dynamic 3D scenes. However, significant gaps remain between dynamic 3DGS and a fully realized 3D Gaussian-based volumetric video. Previous attempts have fallen short in three key areas: (i) excessive model sizes, (ii) limited video duration, and (iii) lack of mechanisms to handle content deviation across extended time spans. All three are crucial for streaming volumetric video from server to client.

To address these challenges, we propose a novel paradigm representing volumetric video as a dynamic 3D Gaussian model, **SwinGS**, yet in a streaming-friendly style compared with previous work. The model maintains a active set of Gaussians to render each frame of the volumetric video. It retires a subset of 3D Gaussians from previous frames and introduces new Gaussians for the next frame. Through assigning each Gaussian an explicitly defined lifespan indicating when it joins and leaves the model, the model can be easily adapted to new content in the subsequent frames. This solves the content deviation issue (iii). On the other hand, the training and transmission of a single bulky model are dismembered into training and transmission of consecutive pieces of data that encapsulates per-frame update, making deriving and streaming of arbitrary length videos theoretically viable, hence addressing concerns (i) and (ii).

To facilitate such paradigm, we innovatively propose a sliding-window based continuous training approach. Within the window, Gaussians contributing to later frames will be jointly optimized with Gaussians contributing to earlier frames. In this setup, Gaussians are shared within a constrained local window achieving a compact representation. During sliding window moving from the head to the tail of video sequence, the continuous training approach guarantees that the rendering quality of earlier frames are unaffected during the optimization for later frames, in the meantime largely reduce the required GPU resources that used to hold bulky training frames in previous dynamic 3DGS methods. Specifically, we train the 3D Gaussian model using Stochastic Gradient Langevin Dynamics (SGLD) and Gaussian relocation, as proposed in 3DGS-MCMC [14]. This allows the model to adapt to various contents across different frames, while keeping a constant number of Gaussians throughout training.

Our contributions can be summarized as follows:

- Instead of previous works focusing on short video clips mainly, our work tries to apply 3DGS onto long volumetric video streaming, identifying its unique challenges.
- Targeting on those challenge, we propose **SwinGS**, a new paradigm to represent volumetric video with a per-frame-update 3D Gaussian model. We also designed and developed a sliding-window based continuous training approach to facilitate the model training, as well as convenient streaming and rendering at client side.
- We implement and evaluate the proposed **SwinGS**, demonstrating its feasibility with DyNeRF datasets. We also develop a WebGL-based viewer that enables easy playback of volumetric video hosted on the cloud storage. A live demo is available at https://swingsplat.github.io/demo/. Codebase will be open-sourced on the acceptance of the paper.

The organization of the paper goes as follows: we first provide a preliminary background for 3DGS and conventional volumetric video representation. Then in Section 3, we talk about how we tackles the challenges in the long volumetric video sequence with our proposed method. Section 4 provides a dive in of our implementation from a system perspective. Finally, we evaluate our method in Section 5.

## 2. Background

### 2.1. 3D Gaussian Splatting

3D Gaussian model is an emerging graphic primitives to represent a 3D scene, and 3D Gaussian splatting (3DGS) [13] is the technique rendering a model into 2D images with given camera poses. In a 3D Gaussian model, the scene is represented with a set of Gaussian points parameterized by covariance $\Sigma$, center position $\mu$, color $c$, and opacity $\alpha$. The intensity of a 3D gaussian at a given location $x$ in the 3D space could be defined as:

$$G(x, \mu, \Sigma) = e^{-\frac{1}{2}(x-\mu)^T \Sigma^{-1}(x-\mu)} \quad (1)$$

In practice, $\Sigma$ is decomposed into a rotation matrix $R$ and a scaling matrix $S$, to guarantee semi-definiteness, as shown in Equation 2. Usually, 3D Gaussians could be visualized as ellipsoids in the 3D space, in which case $R$ could be interpreted as the orientation and $S$ as the length of axises of visualized ellipsoid.

$$\Sigma = RSS^T R^T \quad (2)$$

3DGS is a rasterization-based rendering technique. When it comes to rendering, as shown by Figure 1, ray $r$ emitting from camera center to the queried pixel on the rendering image will traverse a subset of 3D Gaussians, and the final color $C$ of that pixel is computed by alpha blending this subset with the order of z-depth from close to far:

$$C(r) = \sum_{i \in N} c_i \alpha'_i \prod_{j=1}^{i-1}(1 - \alpha'_j) \quad (3)$$

Here $\alpha'_i$ is determined by multiplying opacity $\alpha_i$ with the integration of 3D Gaussian $G$'s intensity along the ray. In practice, integration will be performed by sampling from a 2D Gaussian projected from the original 3D Gaussian.

### 2.2. Dynamic 3DGS

Recent works have extended vanilla 3D Gaussian Splatting (3DGS) to dynamic scenes, primarily following two type of approaches: deformation field-based and 4D primitive-based methods.

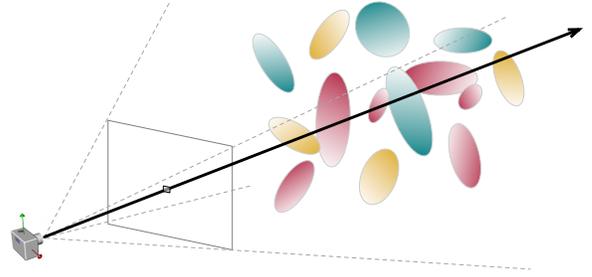

Figure 1: Rasterization process of 3DGS

Deformable 3DGS [38] pioneered the use of multilayer perceptrons (MLPs) to implement deformation fields, allowing 3D Gaussians to exhibit different properties across time frames, following which, [36] enhanced the deformation field's fitting capacity by introducing a hexplane [5] ahead of the MLP. Another early stage work, DynamicGS [23] incorporated rigidity loss to improve tracking accuracy between frames. Several recent works, including SC-GS [10], HiFi4G [12], and Superpoint Gaussian [34], proposed hierarchical structures where higher-layer Gaussians act as deformable skeletons, while lower-layer Gaussians bound to them fit appearance. Taking a different approach, SpacetimeGS [18] used MLPs to encode appearance and parametrized polynomials to represent deformation.

4D primitive-based approaches incorporate time as the fourth dimension of a Gaussian, pioneered by 4DGS [37] and 4D-Roter GS [6]. During rendering, these methods project (also called slice or condition) 4D primitives into 3D Gaussians before processing them through the alpha blending pipeline.

### 2.3. Volumetric Video Streaming

Traditional primitives for streamable volumetric video include point clouds and 3D meshes. Groot [15] presents a system that leverages an advanced octree-based codec to efficiently compress point cloud payloads. Another work, ViVo [9], focusing on point clouds, reduces bandwidth usage by considering visibility based on user viewport. More recently, MetaStream [8] introduced a comprehensive, point cloud-based system for creating, delivering, and rendering volumetric videos in an end-to-end fashion.

Transcoding offers another popular paradigm for volumetric video streaming. Vues [22], for instance, offloads rendering tasks to an edge server instead of rendering received 3D primitives on the user side.

There are some preliminary attempt trying to represent and stream volumetric video with neural radiance field (NeRF) [20, 21]. Yet the use of 3D Gaussians as primitives for volumetric video streaming remains a largely unexplored field, with most dynamic 3DGS research focusing on short video clips lasting less than a dozen seconds. The research most similar to our proposed **SwinGS** is 3DGStream [31], which introduces a Neural Transformation Cache (NTC) as a per-frame deformation field. A recent work, MGA [32] treats network bandwidth as a constraint and optimizes bitrate allocation by tuning various encoding parameters for 3DGS frames to achieve optimal overall rendering quality.

| Primitive | Rendering Quality | Rendering Speed | Accessibility | Storage |
|---|---|---|---|---|
| 3DGS [13] | high | high (>100fps) | medium | medium |
| NeRF [25] | high | low (<10fps) | medium | low |
| point cloud | poor | high | easy | medium |
| mesh | depends | high | poor | medium |

Table 1: Comparison between different 3D primitives for volumetric video

## 3. Motivation

### 3.1. Pitfall of Previous Methods

Among various primitives representing 3D scenes, including NeRF [25], point cloud, and mesh, 3D Gaussian achieves a balance between rendering quality, speed, accessibility, and storage cost, as shown in Table 1. While NeRF delivers high-quality renderings with reasonable storage, its computational demands limit rendering speed. Works including StreamRF [16] and NeRFPlayer [29] attempted to increase rendering speed, but at a cost of dramatically increased model size. Meshes, although relatively compact and rendering-friendly, require extensive manual labor to create, especially for high-quality volumetric videos. Point clouds offer decent frame rates and are easily accessible from 3D dense reconstruction, but their rendering quality correlates with point count, leading to high storage costs and challenges like hole filling.

Recent research on dynamic Gaussian splatting [6, 10, 12, 18, 23, 31, 34, 36–38] demonstrates the potential of 3D Gaussian models for representing 3D scenes. However, these approaches remain incompatible with volumetric video streaming due to several limitations:

**Excessive Model Size**: A naive approach could be constructing static 3D Gaussian models for each frame, which leads to substantial traffic overhead. To tackling with this, one of the baselines, 3DGStream [31], employs per-frame Neural Transformation Codes (NTC) to transform Gaussians between frames. However, this still incurs a storage overhead of approximately 7.8MB/frame, requiring a minimum bandwidth of 200MB/s for 30fps video. Other works [10, 12, 34, 36, 38] utilize a shared neural network for Gaussian deformation or appearance encoding, along with initial Gaussians for the first frame. While this reduces storage costs, the entire model still occupies hundreds of megabytes [18] and must be transmitted at once. Any packet loss during transmission could block the entire rendering pipeline, making it unsuitable for streaming applications.

**Limited Video Length**: One barrier for scaling up the video length is the linearly increased GPU resource requirement. Previous deformation based methods [18, 36, 38] trained the whole set of Gaussian with a random sampling dataloader on the whole set of training views which requires a great amount of GPU memory to preload them. Further, longer video sequence typically requires more Gaussians which also takes up a large fraction of GPU resources. Another barrier is the limited capacity of a single neural network responsible for Gaussian deformation. [18] confirms that increasing clip length from 50 to 300 frames results in a PSNR drop from 29.48 to 29.17 for the Flame Salmon dataset of DyNeRF [17]. Scenes with more substantial motion and longer duration are likely to experience more severe degradation, as the neural network must learn and remember increasingly diverse and distinct Gaussian deformations across frames.

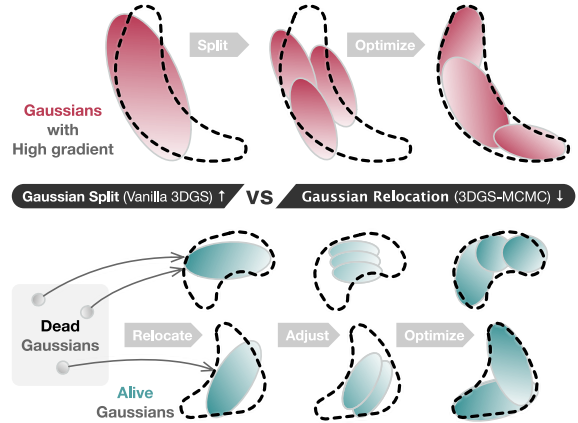

Figure 2: Different Gaussian densification methods (dotted line referring to the shape to fit)

**Long-term Content Deviation**: Most previous methods lack mechanisms for introducing new Gaussians when objects enter the scene or removing them when objects exit. This limitation not only reduces rendering quality but also exacerbates the model size issue, as all Gaussians must be transmitted simultaneously. While fragmentation with multiple dynamic Gaussian models could potentially address this concern, it introduces new challenges such as visual discontinuities between fragments and does not alleviate the burst nature of model transmission, in addition to extra storage cost coming with multiple deformation networks.

### 3.2. 3DGS-MCMC for Bandwidth Shaping

For streamable content, maintaining a uniform data volume across time is crucial. Significant variations in the number of Gaussians between different time frames of a volumetric video can compromise the quality of service (QoS) on the client side, especially when the user is in a high mobility scenario with constrained network bandwidth.

Recent work, 3DGS-MCMC [14], introduces a novel approach to Gaussian densification during model optimization, as shown in Figure 2. Instead of simply splitting one Gaussian into several, 3DGS-MCMC relocates "dead" Gaussians (those with low opacity) to the positions of "alive" Gaussians (those with high opacity hence high presence in the scene). After that, parameters of both "dead" and "alive" Gaussians are adjusted in a way that the Gaussians distribution keeps approximately consistent. With total number of Gaussians keeps constant after densification, this method allows for precise control over the number of Gaussians.

Notably, this relocation operation also naturally facilitates smooth transitions of Gaussian distributions between consecutive frames. Gaussians representing exiting objects are optimized into "dead" Gaussians with diminishing opacity and scale. In subse-quent iterations, these "dead" Gaussians can be repurposed to represent newly appearing objects.

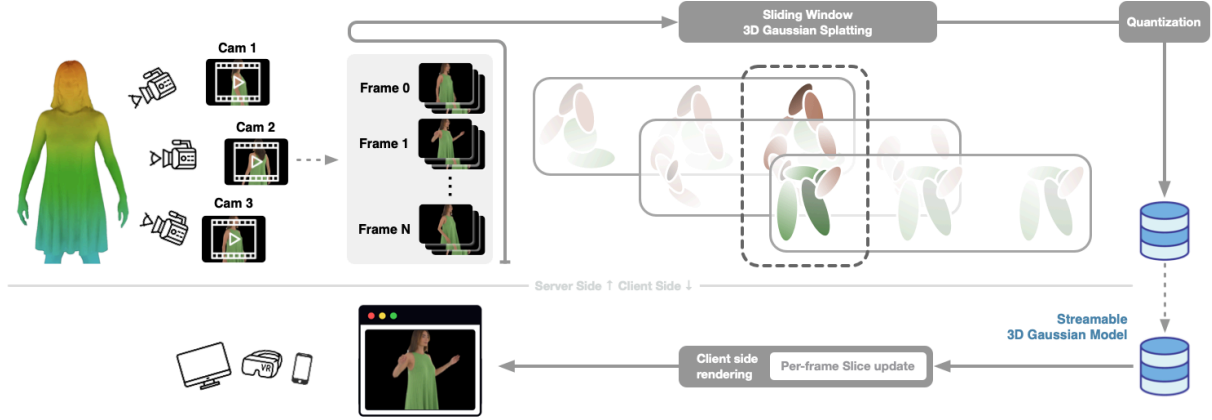

Figure 4: Overview of `SwinGS`

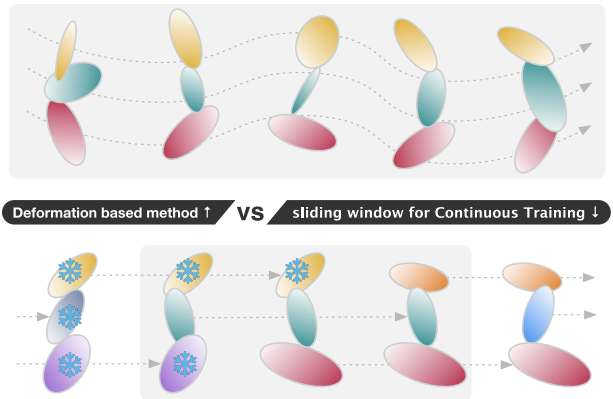

Figure 3: Different paradigm for dynamic 3D Gaussian model (dotted arrow referring to the lifespan of each Gaussian)

### 3.3. Sliding Window for Continuous Training

The key difference between proposed `SwinGS` and previous work is visualized in Figure 3. Previous methods [4, 10, 18, 37, 38] dominantly adopt a deformation-based paradigm where a fixed set of Gaussians in the canonical space is deformed by a carefully designed deformation network at different frame to fitting the dynamics in the video. Yet the limited fitting capacities of deformation network and such tight coupling between frames makes the model hard to train. Hence, complicated modules including hexplane [5] and resfields [24] are integrated to compensate for that, usually with an extra overhead of storage and transmission.

Yet our proposed paradigm assigns a clear lifespan for each Gaussian so that the optimization is always focusing on a short snippet instead of the whole video. We could conveniently optimize those "temporal-local" Gaussians using a sliding window, then derive a streamable model in an incremental style. During the window optimization, Gaussians who also contributes to out-of-window frames are frozen to avoid degradation of previous frames' rendering quality. This combination of per-frame-update paradigm and continuous training approach, makes training on long video sequence feasible and also requires much fewer GPU resources compared with previous methods.

## 4. Design of `SwinGS`

### 4.1. Overview

In this paper, we introduce `SwinGS`, a streaming-friendly paradigm representing volumetric video as a per-frame-update 3D Gaussian model. Figure 4 provides a overview.

`SwinGS` begins with multi-view video input, accompanied by corresponding camera poses. The initial step involves decomposing these videos into individual frames, which are then clustered based on their frame index. This process yields a sequence of folders, each corresponding to a specific frame in the volumetric video.

Following frame clustering, we train a 3D Gaussians model using SGLD and Gaussian relocation to fit various frames within the sliding window. For continuous training, we freeze and archive a small portion of Gaussians at the end of each window training, in the meantime, inject some new optimizable Gaussians for the futural frames. Consequently, we allow each Gaussian to contribute to image rendering across a small spanning of frames.

When it comes to video streaming, similarly, we only need to stream and update the small portion of Gaussians to update the model, which helps substantially reduce bandwidth requirements and makes efficient streaming of volumetric video possible. Yet to further reduce the required transmission bandwidth we quantize Gaussian attributes just like [7, 26] with minimal degradation of the final rendering quality.

The following two subsections will delve into the details of the training process and client rendering respectively.

### 4.2. Continuous Training

The general workflow of model training is outlined in Algorithm 1. The process involves **an outer loop** that shifts a sliding window across the video, iterating from [0, swin_size) to the video's end, and **an inner loop** trains on frames randomly sampled within this sliding window. Here, constant swin_size represents the window length, which also defines the maximum lifespan of a Gaussian and num_gs represents the maximum of Gaussians that will coexist and be active in the model to render an image.

Each of the Gaussian uses two integers to indicates its lifespan: "start" and "expire". The Gaussian will participate into rendering only if the current frame fall into its lifespan. During model training, we have two set of Gaussians in the GPU memory, **gs** and **matured**, as shown in the Algorithm 1 as global variables. To make

**Algorithm 1:** SwinGS Continuous Training

```
1  global gs, matured, stream, trainset
2  const swin_size, num_gs, relocate_period, iterations
3  procedure train_swin(st, ed):
4      for iter in range(iterations):
5          gt = sample_frame_between(trainset, st, ed)
6          frame = ground_truth.frame
7          active_idx = gs.filter(start ≤ frame < expire)
8          active_ma_idx = matured.filter(start ≤ frame < expire)
9          active_gs = gs[active_idx] + matured[active_ma_idx]
10         pred = render(gt.cam, active_gs)
11         loss = loss_func(gt.image, pred) + reg(active_gs)
12         loss.backward()
13         with no_grad:
14             gs[active_idx].param += λ_noise * ε
15             if iter % relocate_period == 0 then
16                 relocate(gs[active_idx])
17 procedure mature(st):
18     mature_idx = gs.filter(start < st)
19     stream.write(gs[mature_idx].detach())
20     matured += gs[mature_idx].detach()
21     matured = matured[-num_gs:]
22     gs[mature_idx].birth = gs[mature_idx].expire
23     gs[mature_idx].start = gs[mature_idx].expire
24     gs[mature_idx].expire = gs[mature_idx].start + swin_size
25 procedure main:
26     gs[num_gs].param = random()
27     gs[num_gs].birth = 0
28     gs[num_gs].start = 0
29     gs[num_gs].expire = swin_size
30     train_swin(0, swin_size)
31     schedule_expire(gs)
32     for st in range(1, trainset.total_frames):
33         mature(st)
34         train_swin(st, st + swin_size)
35     mature(trainset.total_frames)
```

**Algorithm 2:** SwinGS Streaming and Rendering

```
1  global frame, buffer, events, stream, user
2  const swin_size, num_gs, FPS
3  procedure render_thread:
4      while true:
5          active_idx = buffer.filter(start ≤ frame < expire)
6          render(user.cam, buffer[active_idx])
7  procedure update_thread:
8      while true:
9          sleep(1000/FPS)
10         for (target_frame, slice, update) in events:
11             if target_frame == frame then
12                 update_first = slice * slice_size
13                 update_last = (slice+1) * slice_size
14                 buffer[update_first:update_last] = update
15                 break
16         frame += 1
17 procedure main:
18     frame = 0
19     buffer[:num_gs] = stream.read(num_gs)
20     slice_size = num_gs / swin_size
21     threading render_thread
22     threading update_thread
23     while true:
24         update = stream.read(slice_size)
25         target_frame = update[0].birth
26         slice = target_frame % swin_size
27         events.append([target_frame, slice, update])
```

Upon completing training within a sliding window, we increment both the start and end frames by one. We then check if any Gaussian's lifespan begins earlier than the window's start frame. If so, we mature that Gaussian and save its parameters.

**4.2.1. Slice as the optimization granularity.** To further accelerate this process, we evenly divide the entire 3D Gaussians model into several smaller slices. Gaussians within the same slice share identical lifespans and mature together, while different slices have misaligned lifespans. This arrangement ensures that exactly one slice matures in each frame.

Figure 5 illustrates this slicing approach, with `swin_size` set to 5. Each bar represents a group of Gaussians with identical lifespans. Within a slice, multiple bars are positioned in a bumper-to-bumper style across different frames, as older Gaussians trained for previous frames mature and new optimizable Gaussians are introduced. Darker bars denote matured Gaussians, while white bars represent optimizable Gaussians. For training frame #N+1 within the current sliding window, matured Gaussians in slices #0, #1, #4, and optimizable Gaussians in slices #2 and #3 participate in image rendering. After model training within the [N, N+5) window, the sketchy bar of slice #2 will mature, because there will be no further chance to optimize model on frame #N.

### 4.3. Real Time Streaming and Rendering

Algorithm 2 illustrates the rendering process of the 3D Gaussian model on the client's device. At beginning, the first `swin_size` slices consisting of total `num_gs` 3D Gaussians is streamed from the remote server to the client device. Then it is loaded to GPU to render the first frame.

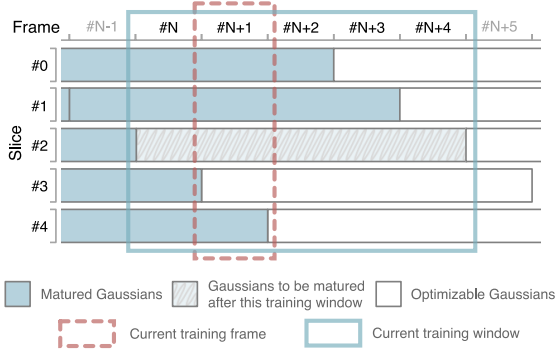

Figure 5: Continuous training with sliding window

it simple, **gs** are those Gaussians we are currently optimizing, usually helping fitting the content in the new frames, while **matured** are those Gaussians that has been snapshotted already and are not optimizable, contributing to the rendering of previous frames as well as current frames. Whenever there is a frame training, both **gs** and **matured** will be used to rendering the image and derive photometric loss yet there is no gradient for **matured** Gaussians.

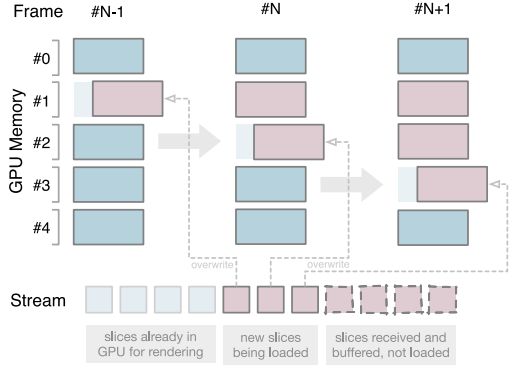

Figure 6: Per-frame slice update in client side

Subsequent slices will be received, processed, and buffered by the client. Those slices will be inserted into the GPU memory replacing expired slices. The `slice_size` in Algorithm 2 represents the number of Gaussians in each slice, while `buffer` abstracts the GPU memory. The GPU memory is divided into `swin_size` slices, with each slice containing `slice_size` Gaussians.

Figure 6 depicts how the client device's GPU memory is updated, in a slice by slice manner, as new streaming data from the server is received. For frame#N-1 slice #1 gets updated, while slice #2 and #3 get updated at frame#N and frame#N+1 receptively.

To decouple GPU rendering from streaming data receiving, which is IO intensive, two dedicated threads are instantiated. Received slice will first be buffered in CPU memory, specifically in the `events` queue. When it is the time to rendering its corresponding frame, `update_thread` will migrate the slice from CPU to GPU memory. `render_thread` keeps rendering images at the same FPS as vanilla 3DGS with current Gaussians in the GPU.

Compared to 3DGStream [31], which recalculates and refreshes all Gaussians for every frame, our approach significantly reduces both streaming traffic and GPU operations.

### 4.4. Implementation

Our demo is built upon the foundation of 3DGS-MCMC [14]. We extended the original codebase by transitioning from a per-frame training approach to a per-window training strategy, as proposed in Algorithm 1. We refactored the `GaussianModel` and `Scene` classes to accommodate Gaussians. A new `SwinManager` class was implemented to handle sliding window. To incorporate the perturbation required by Stochastic Gradient Langevin Dynamics (SGLD) [35], we introduced scaled noise for optimizable active Gaussians post-training for each frame. Our loss function adheres to the practice established in [14], encompassing image quality measurements and regularization terms for opacity $\alpha$ and scaling $S$.

### 4.5. Other Challenges

Adapting 3DGS-MCMC, originally designed for static 3D reconstruction, to our cross-frame Gaussians presented significant challenges. Follows are two primary challenges.

**4.5.1. Adaptive gradient scaling for lr decay.** During the training procedure of vanilla 3DGS[13], learning rate for Gaussians' means is decayed along the iterations to achieves a coarse to fine optimization. Yet in our setup, Gaussians that have been trained for different number of iterations are optimized together. This makes it harder to optimize each Gaussian and decay their learning rate respectively considering Gaussians' means is a single optimizable parameter from the perspective of the optimizer.

As an alternative, we downscale the gradient of each Gaussians' means in accordance to how many sliding windows they have been trained for. It is mathematically equivalent to learning rate downscaling for a SGD optimizer yet a cheap approximate for Adam, which has been proved to be effective in our practice.

**4.5.2. Dynamic dataloader for training set.** The second major challenge stemmed from the limited GPU memory available in our testing environment. The original 3DGS codebase loads all training set images into GPU memory at the initiation of model training. This approach becomes infeasible when dealing with volumetric video containing numerous image sequences from different cameras, each comprising hundreds or thousands of frames. To address this constraint, we applied two key modifications:

- We developed a `LazyCamera` class to load images in a lazy manner, significantly reducing initial memory requirements and improving dataset loading speed.
- To manage memory constraints when training with longer sliding window sizes, we maintain a maximum number of frames in GPU memory. This approach involves dynamically unloading and reloading different frames as needed during the training process.

### 4.6. WebGL Viewer

To visually demonstrate the feasibility of our proposed paradigm, we implemented a web-based viewer building upon the open-source project https://github.com/antimatter15/splat, which was originally designed for rendering static 3D Gaussian models in browsers leveraging WebGL API. We extended this viewer to support rendering the streamable models as proposed by **SwinGS**.

Our modified version incorporated the per-frame slice updates as proposed in Algorithm 2. The streamable model is hosted on Huggingface, a popular online platform for sharing and distributing machine learning models. On the client side, whenever a new slice arrives, the rendering set of Gaussians will be scheduled to be updated with that new slice. However, due to the limitations of web applications in directly manipulating GPU memory, we perform slice updates on the Gaussians buffer within the rendering worker.

## 5. Evaluation

We comprehensively evaluate **SwinGS** using DyNeRF [17] following common practice of previous works. For the metrics, we evaluate **SwinGS** across multiple dimensions, including rendering quality, speed, and traffic cost. Implementation details could be checked in Section 8 in the supplementary material.

### 5.1. Overall comparison

We compare **SwinGS** with previous works that utilize neural rendering to reconstruct dynamic 3D scenes, as shown in Table 2. We evaluated each scenes from the DyNeRF [17] dataset for 300 frames as benchmarks. We set `swin_size=5` with `num_gs=200K`, updating 40K Gaussians per frame.

**SwinGS** surpasses NeRF-based methods like [16] and achieves comparable rendering quality to 3DGStream [31]. While our PSNR is slightly lower than SpacetimeGS [18], **SwinGS** excels in balancing high streamability with low per-frame storage requirements. Unlike baseline methods that require compulsory neural network inference for numerous Gaussians, our approach primarily incurs

| Method | Coffee Martini | Cook Spinach | Cut Beef | Flame Salmon | Flame Steak | Sear Steak | Avg. | FPS | Storage | Long Sequence |
|---|---|---|---|---|---|---|---|---|---|---|
| 4DGaussians [36] | 27.34 | 32.46 | 32.90 | 29.20 | 32.51 | 32.49 | 31.15 | 30 | **0.3MB** | ✗ |
| SpacetimeGS [18] | **28.61** | 33.18 | 33.72 | **29.48** | 33.64 | **33.89** | **32.05** | 140 | 1MB | ✗ |
| StreamRF [16] | 27.84 | 31.59 | 31.81 | 28.26 | 32.24 | 32.36 | 28.26 | 10 | 17.7MB | ✓ |
| 3DGStream [31] | 27.75 | 33.31 | 33.21 | 28.42 | **34.30** | 33.01 | 31.67 | 215 | 7.8MB | ✓ |
| Ours | 27.99 | **33.66** | **34.03** | 28.24 | 32.94 | 33.32 | 31.69 | **300** | 2.1MB | ✓ |
| Ours (quantized) | 27.70 | 33.42 | 33.77 | 28.17 | 32.93 | 32.88 | 31.47 | 300 | 1.2MB | ✓ |

Table 2: Overall comparison with other neural rendering methods
Evaluate on DyNeRF Dataset, PSNR as metrics.

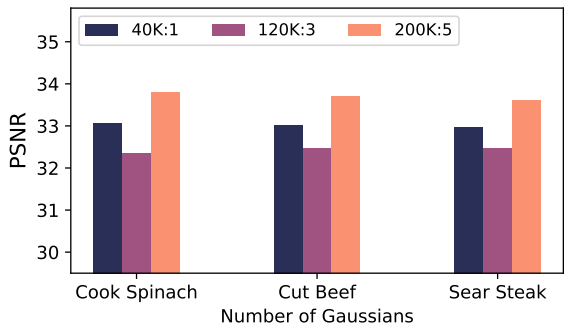

Figure 7: Training with various setup but constraint bandwidth
Label refers to num_gs:swin_size

| Number of Frames | 10 | 30 | 50 | 100 |
|---|---|---|---|---|
| SpacetimeGS[18] | 7.46 | 14.37 | 20.57 | OOM |
| Ours | 2.68 | 2.61 | 2.71 | 2.65 |

Table 3: GPU memory utilization (GB)
when training with different video length costs from GPU memory operations, enabling our model to achieve the best rendering FPS.

### 5.2. Sliding window for constrained bandwidth

The key design parameters in **SwinGS** are swin_size and num_gs. The former determines how many consecutive frames a Gaussian participates in, while the latter defines the total number of Gaussians used to fit a single frame. A larger swin_size typically results in higher degree of content sharing among frames, which saves bandwidth. Conversely, more num_gs can potentially provide better detail in the rendered image, albeit at the cost of increased traffic and storage. The bandwidth required for streaming a model can be formated as:

$$BW = FPS_{video} \times \frac{num\_gs}{swin\_size} \times N_{bytes/GS} \quad (4)$$

This implies that the actual traffic cost is proportional to slice_size = $\frac{num\_gs}{swin\_size}$, given a fixed video FPS and storage cost for each individual Gaussian. Further, Figure 7 demonstrates the effectiveness for sliding window mechanism under a constraint transmission bandwidth. With an appropriate swin_size, our method could boost the rendering quality with shared Gaussian primitives across frames.

### 5.3. Overhead for long volumetric video training

As shown in Figure 3, the key difference between **SwinGS** and previous work is that: in previous work, all the Gaussians of the model will contribute to the rendering of each frame in a deeply coupled way (canonical + deformation), while in **SwinGS**, each Gaussian only binds to frames within a small window. Combined with the dynamic dataloader we proposed in Section 4.5.2, this difference greatly reduce the required GPU resources when training the model, considering at one time, only a small portion of the training views as well as a small subset of all the Gaussians needed to be loaded into GPU memory and contributes to back propagation.

We profile the GPU memory utilization when training the model with different number of total frames in the video sequence. Table 3 shows that, previous method [18] requires a linearly increasing GPU memory while **SwinGS** only requires a fixed amount of resources. Profiling is performed in a NVIDIA 4090 GPU. This video length invariant resource consumption is crucial when we want to train our model for a volumetric video with arbitrary length.

### 5.4. Efficiency for client side rendering

The primary cost introduced by **SwinGS**, compared to vanilla 3DGS, is the memory operation to replace expired Gaussians with new ones in the GPU. This allows for a rendering speed of over 300 FPS when we use the vanilla differentiable Gaussian rasterization module [13] to render the 3D scene.

However, when streaming volumetric video with our WebGL viewer, the scenario becomes a little bit complex. Typically, after chunks of raw byte streams are read from the cloud host, several steps are required before rendering the streaming data into images: raw data preprocessing, depth sorting, and texture generation. The first step involves parsing the binary raw data into Gaussian objects. Depth sorting arranges all Gaussians according to their distance from the camera center, from close to far. The third step, texture generation, is necessary because 3D Gaussians cannot be directly rendered by WebGL; they must be converted into texture data before being delivered to the shader for rendering. All steps introduces additional overhead to the rendering latency.

Figure 8 visualizes the latency composition when rendering volumetric video on a laptop. Raw data processing generally takes less than 1 ms, while depth sorting and texture generation take longer, ranging from 5ms to 18ms. There is a clear correlation between the number of Gaussians and the time required for sorting and texture generation. This is reasonable considering our WebGL implementation does not manipulate the GPU directly and retransmits the texture data corresponding to all active Gaussians to the shader as a whole for every new frame. When configured with swin_size as 4 and num_gs as 200K, it takes approximately 34ms to complete the full pipeline for one frame, resulting in a worst-case video frame rate of around 30 FPS for serialized computation. Considering the three stages could be fully paralleled among video frames, we expect an optimized version achieving over 60fps for video frame rate with texture generation as bottleneck stage for 18ms for our WebGL viewer.

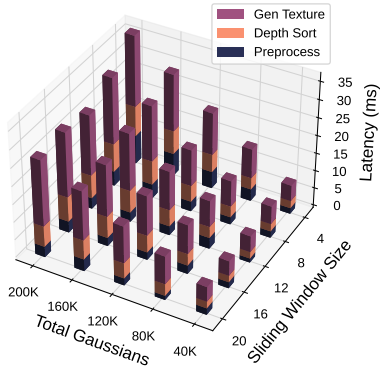

Figure 8: Rendering latency decomposition for WebGL viewer

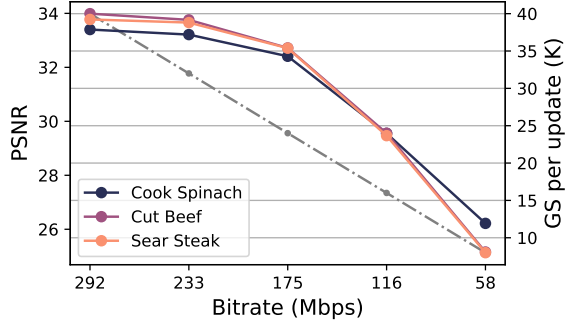

Figure 9: Rate and distortion tradeoff through subsampling Gaussians per slice update

| Stages | preproc | sort | texture | overall |
|---|---|---|---|---|
| MacBook pro (M3 pro) | 3.00 | 5.81 | 18.46 | 27.27 |
| iPhone (A18 pro) | 6.00 | 4.44 | 19.90 | 30.34 |
| iPad (M1) | 6.48 | 4.94 | 22.29 | 33.71 |
| Pixel (Snapdragon 765) | 13.02 | 17.37 | 49.59 | 79.98 |

Table 4: Rendering latency decomposition on different devices (ms)

| Scene | Cook Spinach | Cut Beef | Sear Steak |
|---|---|---|---|
| w/o gradient scaling | 33.35 | 33.44 | 33.62 |
| w/ gradient scaling | 33.81 | 33.71 | 33.62 |

Table 5: Ablation study for adaptive gradient scaling

| Attr | means | rotation | scale | feat | opacity | none |
|---|---|---|---|---|---|---|
| Quant | fp16 | uint8 | fp16 | fp16 | uint8 | - |
| PSNR | 33.74 | 33.51 | 33.81 | 31.21 | 33.63 | 33.81 |

Table 6: Ablation study for Gaussian attributes quantization

Table 4 further profiles the WebGL viewer on a wide range of mobile devices from performance laptop to portable smartphones. Most modern devices are capable of video playback with 30fps.

### 5.5. Adaptive Bitrate Streaming

num_gs and swin_size are controllable parameters that directly impact both bandwidth usage and rendering quality. This characteristic enables adaptive bitrate control, facilitating a smooth streaming experience for users.

We have implemented a naive ABR demo by simply tail-dropping the lowest opacity Gaussians from current active Gaussians set, to fulfill the bandwidth constraint of a simulated poor network. Figure 9 shows the rate and distortion trade off in terms of required bitrate and PSNR. In the real world scenario, during network congestion, we can directly reduce the number of Gaussians per update by transmitting only a subset of Gaussians sampled from each slice with rest marked as empty padding. This approach helps maintaining acceptable video quality under constrained bandwidth.

### 5.6. Ablation Study

#### 5.6.1. Adaptive gradient scaling.

As shown in Table 5, applying adaptive gradient scaling as an approximate of learning rate decay yield a better rendering quality.

#### 5.6.2. Quantization of Gaussian attributes.

We have tried to quantize different Gaussian attributes with a variety of precisions as shown in Table 6. We found that, quantization over Gaussian means, scales, and opacity will not make a big difference towards photometrics, while a subtle degradation has been observed for rotation quantization. Further, spherical harmonics are rather sensitive attributes, so our practice is that we only retains the dc of spherical harmonics and keeps full precision.

### 6. Conclusion

Drawing inspiration from recent advancements in neural rendering, our work, **SwinGS**, adapts 3D Gaussian Splatting (3DGS) techniques to the challenging domain of long volumetric video streaming. We first identify the unique challenges inherent in this task compared to previous dynamic 3DGS task. In response, we propose a novel method that employs a sliding window technique for continuously training 3D Gaussian models and captures Gaussian snapshots for each frame in a slice-by-slice manner.

Our work represents a significant step forward in the realm of volumetric video streaming, leveraging the strengths of 3DGS: compact representation, high rendering quality, and rapid rendering speed. We believe **SwinGS** opens up new avenues for research and development in this exciting field. As the demand for immersive and interactive visual experiences continues to grow, we anticipate that our contribution will catalyze further innovations in real-time volumetric video streaming.

### Acknowledgements

Our **SwinGS** WebGL viewer is built on the basis of Kevin Kwok's https://antimatter15.com/splat/

# Supplementary Material

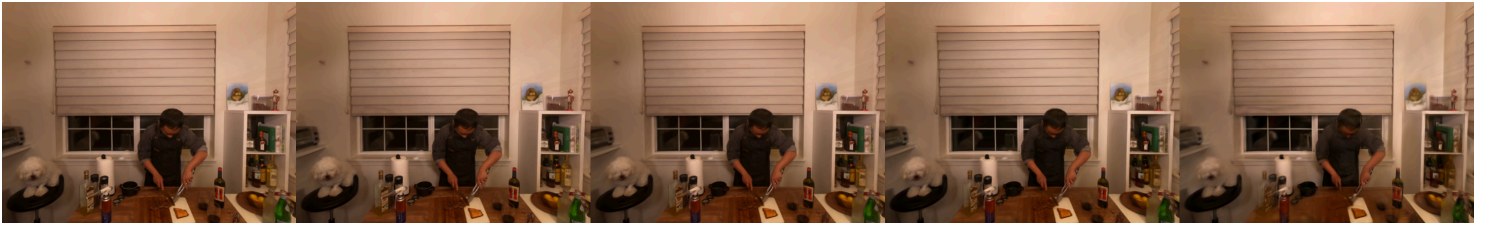

Figure 10: Rate distortion tradeoff for Cut Roasted Beef
From left to right: 100%, 80%, 60%, 40%, 20% subsampling per frame update

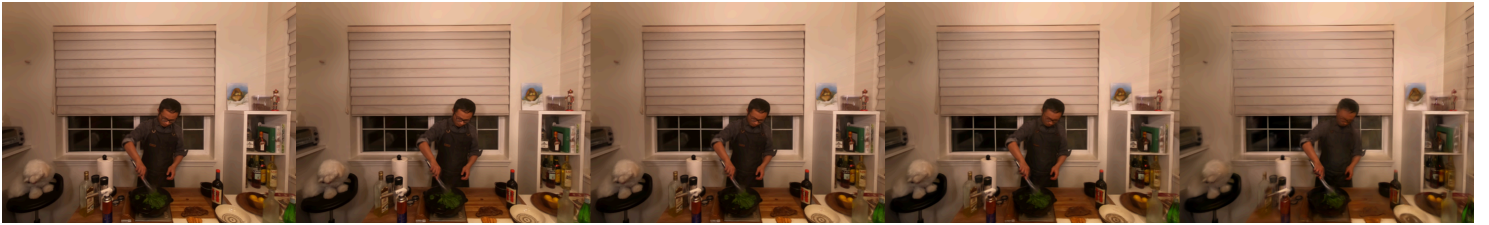

Figure 11: Rate distortion tradeoff for Cook Spinach
From left to right: 100%, 80%, 60%, 40%, 20% subsampling per frame update

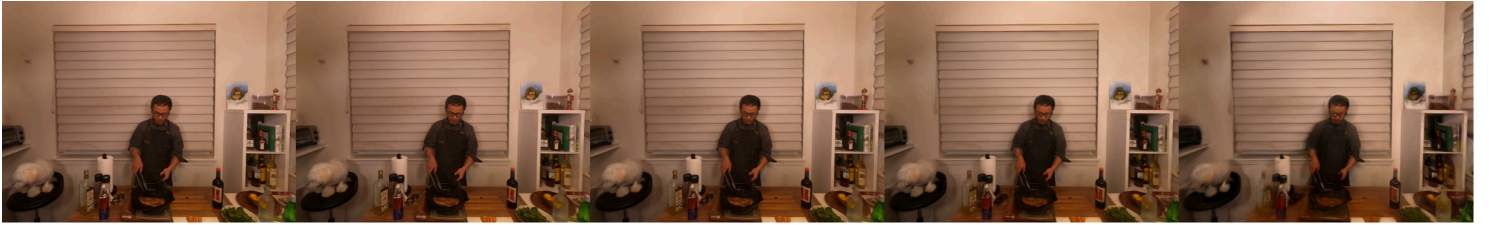

Figure 12: Rate distortion tradeoff for Sear Steak
From left to right: 100%, 80%, 60%, 40%, 20% subsampling per frame update

## 7. Qualitative result for ABR

## 8. Implementation details

### 8.1. Setup

For model training in **SwinGS**, we carefully tuned the hyperparameters to achieve optimal performance. The key parameters were set as follows: `scale_reg` at 1e-2, `opacity_reg` at 2e-2, and `noise_lr` at 5e4. The degree of sphere harmonics function for viewpoint dependent coloring is set to 1 to reduce storage cost. These values were determined through extensive experimentation to balance model accuracy and computational efficiency.

We initialize our model using SfM points derived from COLMAP [28]. This approach provides a strong initial geometry estimate, significantly improving convergence speed and final model quality. We also explored random initialization, but found it prone to overfitting in our experimental setup. By default, our training setup is adapted to the specific characteristics of each dataset as follows:

| Scene | Genesis Iters | Iters | Total Num | Swin Size |
|---|---|---|---|---|
| DyNeRF [17] | 30K | 2K | 200K | 5 |

`Geneiss iters` refers to the training iterations for the very first sliding window, i.e. [0, `swin_size`) and `Iters` refers to training iterations for the following sliding windows. `Total Number` indicates the maximum active Gaussians for a frame.

The choice of Gaussian counts was made to balance model expressiveness with computational efficiency as well as storage cost. DyNeRF scenes benefit from a higher Gaussian count due to the presence of both dynamic foreground elements and static background details. The sliding window size of 5 frames was selected as an optimal trade-off between temporal coherence and computational resources.

## 9. Further discussion

### 9.1. 3DGS and Point Cloud-Based Methods

While Table 1 presents 3D Gaussian Splatting (3DGS) and point cloud methods as alternatives, 3D Gaussians can actually be viewed as an extension of 3D points. In addition to the position (xyz) and color (rgba) properties inherent to points, 3D Gaussians incorporate rotation (R) and scaling (S). This relationship allows for seamless adaptation of point cloud-based volumetric video streaming innovations to the 3D Gaussian representation.

For instance, GROOT [15] employs octrees for efficient geometry data compression, a data structure recently also has been applied to 3D Gaussians in Octree-GS [27] to reduce the total number of Gaussians in a rendering scene. Similarly, RTGS [19] adopts an approach analogous to ViVo [9], utilizing user camera poses to optimize rendering resource allocation across different scene sections.

On the other hand, current off-the-shelf point cloud codec, MPEG Point Cloud Compression (PCC) [11] or a general data compressor like arithmetic entropy encoding could also be integrated to further reduce the transmission load during video streaming, at a cost of longer decoding time on the client side.

We anticipate further extensions of point cloud techniques to 3DGS, for example point cloud super-resolution [39], which could potentially reduce data streaming traffic costs in future implementations.

### 9.2. Limitations and Future Work

As the first effort to adapt 3DGS from short 3D dynamic scenes to long volumetric videos, `SwinGS` faces several challenges that warrant further investigation.

**9.2.1. Inflexible Maturation Schedule.** Current Gaussian maturation process follows a fixed schedule, where Gaussians mature after exactly `swin_size` frames to facilitate efficient batch operations in GPU memory. Although the Gaussian relocation mechanism and "birth" field allow some inter-slice migration, we have yet to fully optimize bandwidth usage with the most informative content. An ideal scenario would involve replacing the least useful Gaussians with the most informative new ones, rather than updating a preassigned fixed subset. For instance, within 40K Gaussians per frame, we could prioritize updates for Gaussians representing human motion while less frequently updating those depicting static backgrounds.